\documentclass{article}
\newcommand{\cst}{\ensuremath{c_{\rm ST}}}
\newcommand{\cem}{\ensuremath{c_{\rm EM}}}
\title{A tight constraint on some varying speed of light theories}

\author{Ahmad Shariati$^1$ and Nosratollah Jafari$^2$
\\[5pt] $^1$ \textit{Department of Physics, Alzahra University,}
\\[0pt] \textit{Tehran 19938-91167, Iran.}\\ \texttt{\normalsize E-mail: shariati@mailaps.org}
\\[5pt] $^2$ \textit{Institute for Advanced Studies in Basic Sciences,}
\\[0pt] \textit{P.O. Box 1159, Zanjan 45195, Iran}\\ \texttt{\normalsize E-mail: njafary@iasbs.ac.ir}
} 
\date{Feb 2005}
\begin{document}
\maketitle
\begin{abstract} We argue that if the velocity of electromagnetic
waves in vacuum, \cem, is different from the limiting velocity for
massive particles, \cst, then, a proton moving with velocity $\cem
< v < \cst$ would radiate Cherenkov radiation.  Because we have
seen protons with energy $\sim 10^{19}$~eV coming from a distance
of order more than $1$~Mpc, we can put a constraint on $\Delta :=
(\cst - \cem)/\cst$: $\Delta < 10^{-27}$!
\end{abstract}
\par\noindent In the past several years,
some people argued that a varying speed of light may be a solution
to some puzzles in cosmology [1]. But the speed of light being a
universal constant lies at the heart of theories of spacetime, and
one should clarify exactly what is the meaning of a varying speed
of light. As Ellis and Uzan pointed out in \cite{EU}, there are
four different quantities to be named ``velocity of light, $c$''.
One is the constant appearing in the Lorentz transformations, in
the metric $\eta_{\mu\nu}$ of special relativity, and in the
$g_{\mu\nu}$ of general relativity. This is the limiting speed of
massive particles and observers. Following their notation, we name
it \cst.  Another notion of the speed of light, is the velocity of
the propagation of the electromagnetic waves.  We name it \cem.
\par In general relativity, and in the standard model, $\cst=\cem$,
but logically they might differ.  Let $\delta c = \cst - \cem$,
and $\Delta = \delta c /\cst$.
In this letter, we argue that the observation of ultra high energy
cosmic rays (protons), put a higher limit on $\Delta$: $\Delta < 10^{-27}$!

\par The argument goes as follows.  $\cem = 1/\sqrt{\epsilon_0\, \mu_0}$,
where $\epsilon_0$ and $\mu_0$ are the permittivity and
permeability of vacuum. \cem\ cannot be greater than \cst, because
it leads to a violation of the special relativity.  If $\cem <
\cst$, then consider a proton moving with velocity $\cem < v
<\cst$.   From the theory of Cherenkov radiation \cite{Jackson} we
know that if a charged particle moves in a medium faster than the
speed of electromagnetic waves in that medium, then
\begin{enumerate}
\item it emits electromagnetic waves in those frequencies for which
$ v > \cem(\omega)$,
\item the energy loss, per unit length, is given by
$$ \frac{dE}{dx} = - \frac{e^2\, \mu_0}{4\, \pi} \int \left( 1 - \frac{\cem^2(\omega)}{v^2} \right)
\, \omega \, d\omega $$ the integral begin over all frequencies
satisfying $v > \cem(\omega)$.
\end{enumerate}
In vacuum, \cem\ does not depend on frequency, so that
$$ \left\vert\frac{dE}{dx}\right\vert =  \frac{e^2\, \mu_0}{4\, \pi} \left( 1 - \frac{\cem^2}{v^2} \right)
\int_{0}^{\omega_{{\rm max}}} \omega \, d\omega $$
where $\omega_{{\rm max}}$ is a cutoff frequency.
\par Now, we know that protons with energy $\sim 10^{19}$~eV, i.e., $\sim 1$~J, reach the
Earth \cite{cosmic1,cosmic2}. The velocity of these protons is, to
a very good approximation, \cst. Therefore,
$$ 1 - \frac{\cem^2}{v^2} \simeq 1 - \frac{\cem^2}{\cst^2} \simeq 2\, \Delta. $$
So that,
$$ \left\vert \frac{dE}{dx}\right\vert \simeq
2\, \Delta \, \frac{e^2\, \mu_0}{4\, \pi} \, \frac{1}{2} \,
\omega_{{\rm max}}^2. $$ And, we know that these protons have come
from a distance of the order of $1$~Mpc, or $10^{22}$~m
\cite{cosmic1,cosmic3}.  Therefore, $\left\vert dE/dx\right\vert <
10^{-22}\, {\rm J}\,{\rm m}^{-1}$. Setting $e\sim 10^{-19}$~C,
$\mu_0 = 4\, \pi\times 10^{-7}$, and the reasonable value
$10^{25}\, {\rm s}^{-1}$ for $\omega_{{\rm max}}$, corresponding
to photons with energy $\sim 10^{10}$~eV, we get
$$ \Delta < 10^{-27}. $$


\begin{thebibliography}{99}
\bibitem{Mag} J. Magueijo, Rept. Prog. Phys. {\bf66} (2003) 2025.
\bibitem{EU} G.\ F.R.\ Ellis, J.-P. Uzan, ``~`$c$' is the speed of light, isn't it?'',
arXive: gr-qc /0305099
\bibitem{Jackson} J.\ D.\ Jackson, {\it Classical Electrodynamics}, $3^{{\rm ed}}$ edition,
Wiley, 1999, p. 638.
\bibitem{cosmic1}  M.\ Nagano, A.\ A.\ Watson, Rev. Mod. Phys. {\bf72} (2000) 689.
\bibitem{cosmic2} R.\ U.\ Abbasi {\it et al.}, Phys. Rev. Lett. {\bf92} (2004) 151101.
\bibitem{cosmic3} S. Sarkar, Ultra-high energy cosmic rays and new
 physics, arXive: hep-ph/0202013.; M.\ Lemoine, Ultra-high energy
 cosmic ray propagation in the Universe, arXive: astro-ph/0501124.
\end{thebibliography}
\end{document}